# CO Line Emission and Absorption from the HL Tau Disk – Where is all the dust? [1]

Sean D. Brittain
National Optical Astronomy Observatory
Tucson, AZ 85742
brittain@noao.edu

Terrence W. Rettig
Center for Astrophysics
University of Notre Dame
Notre Dame, IN 46556

Theodore Simon
Institute for Astronomy
University of Hawaii
2680 Woodlawn Drive
Honolulu, HI 96822

Craig Kulesa
University of Arizona
Steward Observatory
933 North Cherry Ave.
Tucson, AZ 85721

*Submitted February 17, 2004*

Abstract

We present high-resolution infrared spectra of HL Tau, a heavily embedded young star. The spectra exhibit broad emission lines of $^{12}$CO gas phase molecules as well as narrow absorption lines of $^{12}$CO, $^{13}$CO, and C$^{18}$O. The broad emission lines of vibrationally-excited $^{12}$CO are dominated by the hot (T~1500 K) inner-disk. The narrow absorption lines of CO are found to originate from the circumstellar gas at a temperature of ~100 K. The $^{12}$CO column density for this cooler material ($7.5 \pm 0.2 \times 10^{18}$ cm$^{-2}$) indicates a large column of absorbing gas along the line of sight. In dense interstellar clouds, this column density of CO gas is associated with $A_V \sim 52$ magnitudes. However, the extinction toward this source ($A_V \sim 23$) suggests that there is less dust along the line of sight than inferred from the CO absorption data. We discuss three possibilities for the apparent paucity of dust along the line of sight through the flared disk: 1) the dust extinction has been underestimated due to differences in circumstellar grain properties, such as grain agglomeration; 2) the effect of scattering has been underestimated and the actual extinction is much higher; or (3) the line of sight through the disk is probing a gas-rich, dust-depleted region, possibly due to the stratification of gas and dust in a pre-planetary disk. Through the analysis of hot ro-vibrational $^{12}$CO line emission, we place strong constraints on grain growth and thermal infrared dust opacity, and separately constrain the enhancement of carbon bearing species in the neighboring molecular envelope. The physical stratification of gas and dust in the HL Tau disk remains a viable explanation for the anomalous gas to dust ratio seen in this system. The measured radial velocity dispersion in the outer disk is consistent with the thermal linewidths of the absorption lines, leaving only a small turbulent component to provide gas-dust mixing.

[1] The data presented herein were obtained at the W. M. Keck Observatory, which is operated as a scientific partnership among the California Institute of Technology, the University of California and the National Aeronautics and Space Administration. The Observatory was made possible by the generous financial support of the W.M. Keck Foundation.

Subject headings: circumstellar matter --- stars: formation --- stars: individual (HL Tau) stars: pre-main-sequence --- planetary systems: formation---planetary systems: protoplanetary disks

1. INTRODUCTION

HL Tau is a nearby (d=140 pc) low-mass pre-main-sequence star (Elias 1978). Its high foreground obscuration and large submillimeter continuum flux indicate that a massive circumstellar disk surrounds the nascent star. $^{13}$CO spectral line interferometry of HL Tau reveals circumstellar material with a radius of 1400-2000 AU, a mass of 0.03-0.14 $M_\odot$, and a mass accretion rate of $\sim 10^{-5} M_\odot$ yr$^{-1}$ (Sargent & Beckwith 1991; Hayashi, Ohashi, & Miyama 1993). The highest resolution interferometric maps reveal a dense, compact inner torus ($r \sim$60-180 AU) with a mass of $\sim 0.05\ M_\odot$ (Lay et al. 1994; Mundy et al. 1996; Wilner et al. 1996). These properties make the HL Tau system particularly important because it is considered one of the best, nearest prototypes for the Solar Nebula at a very early stage of planet formation (Beckwith et al. 1990, Beckwith & Sargent 1993).

The dusty disk surrounding HL Tau is thought to extinguish most (if not all) of the visible light such that $A_V \sim$ 22-30 (Monin et al. 1989; Beckwith & Birk 1995; Stapelfeldt et al. 1995). Close et al. (1997) have imaged HL Tau at infrared wavelengths with an adaptive optics camera and find that the 2 µm emission of HL Tau is dominated by the unresolved inner edge of the accretion disk, with just 12% of the light being contributed by the stellar photosphere, half the value determined by Greene & Lada (1996). From the infrared color excess, Close et al. (1997) conclude that $A_J$=7.7 magnitudes and $A_V \sim$24 magnitudes from the Rieke & Lebofsky (1985) interstellar reddening law. Lucas et al. (2004) confirm the J-band extinction inferred by Close et al. (1997) using high resolution imaging polarimetry and find the inclination is 66°-77°. In contrast, modeling by Men'shchikov, Henning, & Fischer (1999) envisions the dense torus surrounding the central protostar to be inclined by 43°, with an optical depth of $\tau_V \sim$33, one-third of which is due to gray extinction. If this is correct, then no light from the source is directly observed at any wavelength. However, Whittet et al. (1988, 1989) find minimal silicates, water ice, or CO ice towards HL Tau suggesting that either the extinction is overestimated or the solid state features have been processed considerably.

The capabilities of high-resolution infrared spectroscopy offer new insight into the interpretation of these data. Ro-vibrational emission lines of CO probe the hot, excited gas of circumstellar disks in the dynamic innermost regions of disks with little envelope contamination. Absorption lines of the same species provide accurate measurements of the physical conditions and kinematics in a pencil-beam milliarcsecond column of gas along the line of sight to the star and accreting disk.

In this paper, we present a detailed absorption and emission line analysis of CO toward the circumstellar disk around HL Tau. High-resolution near infrared observations of $^{12}$CO, $^{13}$CO, and C$^{18}$O are depicted in § 2. In § 3 we use the absorption lines to show the column density of gas we observe is much larger than one might expect based on the extinction reported for HL Tau. We attribute the anomalous gas/dust ratio (i.e. missing dust) to either grain agglomeration or an indication of the dynamical stratification of the evolving circumstellar disk. We attempt to constrain these possibilities in § 4 through a careful analysis of the rich $^{12}$CO emission line spectrum. The resulting constraints upon the line opacity and 5 µm optical depth place strong limits on the contribution of "gray dust opacity" to the overall obscuration of HL Tau. This suggests that grain growth is unlikely to be the sole cause of the enhanced gas to dust ratio. We find that the vertical stratification of gas and dust in the inclined circumstellar disk remains a plausible explanation.

## 2. OBSERVATIONS

Spectra of the v = 1-0 ro-vibrational branch of $^{12}CO$, $^{13}CO$, $C^{18}O$, and of the hot band (v = 2-1) CO lines of HL Tau were obtained at a resolving power of 25,000 at the W. M. Keck Observatory, using the NIRSPEC cross-dispersed echelle spectrograph (McLean et al. 1998). Table 1 summarizes the parameters of our observations, which were taken March 23, 2002.

In the data reductions, we used a series of flats and darks to remove systematic effects at each grating setting. The 2-dimensional frames were cleaned of systematically hot and dead pixels as well as cosmic ray hits, and were then resampled spatially and spectrally in order to align the spectral and spatial dimensions along rows and columns, respectively (DiSanti et al. 2001, Brittain et al. 2003). Absolute flux calibration was achieved through observations and reduction of standard stars.

Observations within the M-band are dominated by a significant thermal (~300 K) continuum background, upon which are superimposed night sky emission lines. The intensities of the telluric lines depend not only on the air mass, but also on the column burden of atmospheric water vapor, which can vary both temporally and spatially over the course of the night. In order to cancel most of the atmospheric + background effects, we nodded the telescope by a small distance (typically 15″) along the slit between two positions (A and B), in an {A, B, B, A} sequence. Each step corresponded to 1 minute of integration time. Combining the scans as (A-B-B+A)/2 canceled the telluric features to first order. Subsequently, the FWHM of the spatial profiles in the "A" and "B" rows were extracted to obtain the spectra for both positions. To model the atmospheric transmittance function of the combined spectrum, we used the Spectrum Synthesis Program (SSP, Kunde & Maguire 1974), which accesses the updated HITRAN 2000 molecular database (Rothman et al. 2003). For each grating setting, the optimized model established the column burden of absorbing atmospheric species, the spectral resolving power, and the wavelength calibration. We then divided the atmospheric model (scaled to the observed continuum) into the observation to reveal the ratioed spectrum of HL Tau.

The resulting absorption and emission spectra of HL Tau are plotted in Fig. 1. The individual lines of the different branches and isotopes are identified in each panel of the figure. The observed position, rest position, Doppler shift and equivalent width (W) for the $^{12}CO$ *absorption* lines are listed in Table 2a. Tables 2b and 2c present the same information for the $^{13}CO$ and $C^{18}O$ absorption lines, respectively. The observed position, rest position, Doppler shift, and flux, F, for the CO *emission* lines of HL Tau are given in Table 3.

The low-J emission lines have absorption features superimposed on them (Figure 1). To measure their flux, we fit the emission lines with a Gaussian curve $g(f)=a_0 exp(-(f-a_1)^2/2a_2^2)$, where $a_0$ is the height of the emission feature, $a_1$ its center position, and $a_2$ its width. Since the high-J lines are generally cleaner than the low-J lines, we fit them first in order to determine the Doppler shift and width of the lines. For the low-J emission lines, we mask the portion of the line profile that is affected by the absorption feature, and minimize the residual between the fit and the remainder of the line by varying the term $a_0$. In the fit, $a_1$ is fixed, since we know the rest position of each line as well as its Doppler shift, and we use the average width of the high-J lines to fix $a_2$.

## 3. ANALYSIS OF THE ABSORPTION LINES

### 3.1. *CO Isotopic Column Densities and Rotational Temperatures*

An absorption-line study of HL Tau samples conditions in an extended disk over a very long path length but gains the advantage of the high angular resolution that is defined by the small physical extent of the background source of illumination (generally a stellar diameter in size). Because an entire ro-vibrational branch of CO can be observed simultaneously in the same "pencil beam" absorbing column, many of the systematic uncertainties that compromise lower resolution submillimeter measurements are avoided. CO absorption studies have been used to constrain physical conditions in interstellar clouds, in diffuse gas at UV wavelengths (Roberge et al. 2001), and in dark clouds at infrared wavelengths (Black & Willner 1984, Kulesa & Black 2002, Brittain et al. 2003). In this section we present the analysis of the near-infrared ro-vibrational absorption bands of CO and its rare isotopomers (Figure 1).

We have measured equivalent widths of absorption line profiles from the fully processed, normalized spectra in two ways, first by summing the absorption area, and second via Gaussian fits to the line profiles. Uncertainties were estimated from the mean deviations from the Gaussian fits and by the formal uncertainty in the fitted continuum level.

Even though the fundamental $^{12}CO$ absorption lines are only 40% deep at the 12 km s$^{-1}$ resolution of our spectra, they are highly saturated. This is confirmed by the unambiguous detection of $^{13}CO$ and $C^{18}O$ v=1-0 absorption (in the interstellar medium these isotopomers are ~60 and ~500 times rarer than $^{12}CO$; Wilson & Rood 1994) and the v=2-0 overtone band of $^{12}CO$, which is ~100 times weaker than the v=1-0 band. Because the effects of saturation must be corrected for, we have determined column densities and level populations from a curve of growth (COG) analysis (c.f. Spitzer 1978), which relates the measured equivalent widths to column densities by taking into account the effects of opacity upon a Gaussian line profile. The derived column density for a measured equivalent width only depends upon one parameter, the Doppler broadening of the line, $b$ (= $\sigma_{RMS}/1.665$, where $\sigma_{RMS}$ is the RMS linewidth).

When, as in the case of HL Tau, the absorption lines are unresolved and $b$ cannot be evaluated directly from the observed spectrum, the intrinsic width of saturated lines can be inferred by the following complementary methods:

1) *Comparison of the P and R branch lines* - CO exhibits absorption lines in both *P* (*J"*=*J'*+1) and *R* (*J"*=*J'*-1) branches, which have different oscillator strengths yet probe the same energy levels, e.g., the P(1) absorption line originates from the same J=1 level as the R(1) absorption line. Any differences in the column density derived from lines that share a common level must be due to optical depth, which is related to $b$. The line width, $b$, can be used to determine the optical depth and adjusted so that the derived level populations from the two branches agree as closely as possible.

2) *Comparison of fundamental and overtone lines* – Despite the presence of heavily saturated lines in the $^{12}$CO fundamental band, their equivalent widths should yield comparable column densities to the much weaker *overtone* transitions that arise from the same lower state. For example, both (2,0) R(3) and (1,0) P(3) stem from the (v=0, J=3) level. Again, $b$ can be adjusted to ensure that the level populations derived from the different bands are consistent.

3) *Linearization of the rotational lines* - The (v=0, low-J) transitions are thermalized at densities as low as $n_H \sim 10^{3-4}$ cm$^{-3}$, and at even lower densities due to radiative trapping in the rotational lines with high opacity. Therefore, the low-J lines are the ones most likely to exhibit a thermal population distribution. The line width that best linearizes a plot of the level populations to a common temperature in an excitation diagram is used.

Subject to the above constraints, the best fitting $b$ value in the COG analysis for HL Tau is $1.3 \pm 0.1$ km s$^{-1}$. The consistency of all data to this common velocity dispersion is depicted in the curve of growth diagram of Figure 2. This Doppler parameter corresponds to a FWHM line width of 2 km s$^{-1}$ and defines a low velocity dispersion in the HL Tau disk along the line of sight. With a measurement of $b$, equivalent widths can be directly related to column density. The corresponding column densities of $^{12}$CO, $^{13}$CO and C$^{18}$O are logarithmically plotted as a function of energy above the ground state in Figure 3. Uncertainty in N($^{12}$CO) from the overtone lines, estimated from the measurement of unsaturated lines, is small and dominated by measurement errors in the equivalent widths of the lines. The uncertainty in N(C$^{18}$O) is dominated by noise in the spectra as the C$^{18}$O v=1-0 lines are weak, making accurate equivalent widths difficult to measure. Owing to the higher optical depth in the $^{13}$CO lines, uncertainty in N($^{13}$CO) is dominated by the uncertainty in $b$.

The temperature and total column density of each species is presented in Table 4. Because a linear slope in such an excitation diagram is proportional to T$^{-1}$, the rotational temperature of CO can be determined from satisfactory linear least squares fitting of the excitation diagram. All of the opacity-corrected data are well fit with a single temperature component. As the low-J lines are readily thermalized by collisions with atomic and molecular hydrogen at modest densities of $n_H = 10^{3-4}$ cm$^{-3}$, this rotational temperature closely approximates the actual kinetic temperature of the gas, and relates its approximate location in the circumstellar system. Around HL Tau, kinetic temperatures of ~100K indicate that the gas we observe in absorption is fairly distant (>1 AU) from the central star.

### 3.2 *Isotopic abundance of CO*

We expect the material from the envelope will generally reflect the elemental composition of dense clouds in the interstellar medium. The results in Table 4 show that the isotopic abundance ratio, $^{12}$CO/$^{13}$CO, is 76±9. This value is consistent with the isotopic composition of CO measured in Orion A (Langer & Penzias 1990). This ratio ranges from 67±3 to 79±7 in different parts of the cloud, which they attribute to selective dissociation. (Langer & Penzias 1990). The data in Table 4 also implies that $^{12}CO/C^{18}O$ = 800±200 which is consistent with the composition presented by Langer & Penzias (1990) and within 3σ of the canonical interstellar value of 560±25 (Wilson & Rood, 1994). While these abundances are generally consistent with the range expected from the interstellar medium, the abundance of the rare isotopomers lies toward the low end of the scale. The under-abundance of the rare isotopomers hints that they are suppressed by isotope-selective dissociation via the Lyman-alpha and UV flux from HL Tau and its attendant accretion disk. That is, the surface layers of the disk will be proportionally more abundant in $^{12}CO$ than the rare isotopes, which are too tenuous to provide any measure of self-shielding at such low column densities. This is reflected in the uniformly higher excitation temperatures of the $^{12}CO$ lines, suggesting that the column of $^{12}CO$ is slightly biased towards gas heated by e.g. a stronger radiation field.

### 3.3. *HL Tau's Anomalous Gas/Dust Ratio*

Observations of the gas conditions and gas/dust ratios in the outer disk regions can constrain models of disk evolution and set the stage for models of planet formation. Theoretical models of disks predict that dust preferentially settles to the mid-plane, leaving behind a dust-depleted atmosphere at higher vertical scale heights (Goldreich and Ward 1973; Miyake & Nakagawa 1995; Youdin & Shu 2002; Chiang 2003), but such predictions have not been observationally tested.

Previous estimates of the extinction toward HL Tau suggest $A_v$ ~ 22-30 magnitudes (Monin et al. 1989; Beckwith & Birk 1995; Stapelfeldt et al. 1995, Close et al. 1997). The column density from our CO data implies substantially larger values of $A_V$ for the same line of sight. Our measured total CO column density is $7.5 \pm 0.2 \times 10^{18}$ cm$^{-2}$ (Table 4). To determine the column density of $H_2$ we adopt the $CO/H_2$ abundance ratio in cold, dark clouds ($1.56 \pm 0.12 \times 10^{-4}$) measured by Kulesa (2002). This abundance ratio is also consistent with the $CO/H_2$ ratio in disks calculated by Glassgold, Najita, & Igea (2004). While there are a variety of effects that may deplete CO relative to $H_2$ (e.g. selective dissociation due to the inability for CO to self shield itself as efficiently as $H_2$ and/or the deposition of CO on grains), it is highly unlikely that CO can be enhanced. This ratio implies $N(H) = 2 \times N(H_2) = 9.6 \pm 0.8 \times 10^{22}$ cm$^{-2}$. For a normal interstellar gas-to-dust ratio $A_V/N_H = 5.4 \times 10^{-22}$ mag cm$^{-2}$ (Mathis 1990)[2] this amount of hydrogen corresponds to an optical extinction of $A_V = 52 \pm 4$ magnitudes, which is more than twice the extinction reported by Close et al. (1997). If the CO is depleted due to selective dissociation, then $CO/H_2$ is likely $<1.5 \times 10^{-4}$ and the inferred extinction is even larger. In the extreme, if we were to adopt the abundance ratio appropriate for a translucent cloud (~$10^{-5}$), the result would be an implausibly large extinction of $A_V > 500$.

How can this discrepancy be understood? There are at least three possibilities that may lead to the anomalous gas/dust ratio toward HL Tau: 1) the dust extinction has been underestimated due to differences in circumstellar grain properties, such as grain agglomeration; 2) the effect of scattering has been underestimated and the actual extinction is much higher; or (3) the line of sight through the disk is probing a gas-rich, dust-depleted region, possibly due to the stratification of gas and dust in a pre-planetary disk. Confirmation of the result by Close et al. (1997) comes from high resolution NIR polarimetry data by Lucas & Roche (1998) and Lucas et al. (2004). They find an unpolarized, unresolved point source in the L-band, and their modeling of the high-resolution NIR polarimetry data in J, H, and K imply infrared extinctions that are consistent with Close et al. (1997). Thus it is unlikely the observed extinction has been underestimated due to scattering. In the next section we examine the suggestion that grain agglomeration in the upper disk around HL Tau is removing the small grains that comprise the bulk of UV, optical and infrared extinction.

---

[2] Kulesa (2002) has confirmed this relationship for $A_V/N_H$ in dense clouds with extinctions as high as 60 magnitudes.

## 4. ANALYSIS OF THE HOT $^{12}$CO EMISSION LINES

Dust grains in the upper disk can grow to sizes of a few microns (D'Alessio et al. 2001), and the settling time for these small particles, $\sim 10^7$ years (Miyake & Nakagawa 1995), is comparable to the lifetime of disks around T Tauri stars. For their model of HL Tau, Men'shchikov et al. (1999) argue for a substantial population of large grains that produce ~10 magnitudes of gray extinction. However, settling velocities are size dependent and will quickly remove the larger particles from the disk atmosphere (Nakagawa, Sekiya & Hayashi 1986, D'Alessio et al. 2001). We can observationally quantify the effects of gray extinction by using our CO measurements at ~5 μm to constrain the extinction at M-band and thus the allowed gray extinction

To accomplish this goal, in § 4.1, we use the kinematics and excitation of the vibrationally excited $^{12}$CO emission lines observed at 4.7 μm to constrain the warm gas location in the HL Tau disk. In § 4.2, we calculate the column density of the warm gas in v=2 and use its rotational temperature to extrapolate the column density to all other vibrational energy levels. Once we have the observed column density of gas, we correct for beam dilution (§ 4.3). Then the interstellar extinction at M-band is constrained by requiring the rotational lines in the v=1-2 band be optically thin, which is implied by the linear v=2-1 excitation plot.

### 4.1. *The Origin of the Line-Emitting Gas*

The spatial resolution of our emission-line observations is limited by the atmospheric seeing of ~0.7″ at Keck II (FWHM in M-band), which translates to ~100 AU at the 140 pc distance of Taurus. The emission lines are not extended relative to the point-spread-function of the star. Also the Doppler shift of the emission lines is consistent with that of the absorption lines (48±4 km/s and 47±1 km/s respectively; see Tables 2 and 3), so it is reasonable to infer that the warm gas is fully circumstellar.

Models of young disks predict that the bulk motions of gas and dust around the disk should be within a few percent of Keplerian orbital velocities (Adachi et al.1976), so the inclination of the disk and the width of the Doppler-broadened emission lines constrain the location of the emitting gas. Najita et al. (2003) point out that the emitting area of gas can be determined from the spectrally resolved line profile. The velocity of gas at the inner radius of the emitting region corresponds to velocity half width at zero intensity (HWZI), and the outer radius of the emitting region corresponds to twice the radius inferred from the velocity half width at half maximum (HWHM; Najita et al. 2003). The average HWZI of the v=2-1 lines observed from HL Tau is 90 km s$^{-1}$. If we adopt an inclination of 71±5° (Lucas et al. 2004) and a stellar mass of 0.7 $M_{Solar}$ (Close et al. 1997), then the inner radius is 0.066 AU. The average HWHM of the v=2-1 lines is 45 km s$^{-1}$, which implies that the outer radius of the emitting region is 0.53 AU. Thus the area of the emitting region is 0.86 AU$^2$. Najita et al. (2003) note that the outer radius of the emitting gas inferred from the line profile could be a few times larger if the intensity of the emission falls off with radial distance within the disk. An uncertainty of that size is not critical in the discussion that follows.

### *4.2 Column Density of Hot CO gas*

High-resolution infrared spectroscopy of ro-vibrational transitions of $^{12}$CO(1,0) and $^{12}$CO(2,1) e*mission* lines provide a sensitive probe of hot circumstellar gas. In dense regions, where $n_H \sim 10^{10}$ cm$^{-3}$, CO is most easily thermally excited by collisions (Najita et al. 1996, Carr et al. 2001). Figure 4a demonstrates the v=1-0 emission lines are optically thick. We infer that the v=2-1 hot-band emission lines of CO are optically thin because the v=3-2 lines are not observed (Figure 1) and the excitation plot for this band is linear (Figure 4b). A straight-line fit to the measured rotational line strengths in the Boltzmann plot for the hot band yields a temperature of 1,500±100 K with a reduced $\chi^2$=1.05 (Fig. 4b).

The column density of an optically thin transition is given by

$$N_{v'J'} = 4\pi F_{v'v''J'J''}/(\Omega A_{v'v''J'J''} hc\tilde{v}).$$

Here, F is the measured flux in each line, $\Omega$ is the solid angle subtended by the beam, A is the Einstein emission coefficient, and $\tilde{v}$ is the frequency of the transition in wavenumbers. The symbol, v designates the vibrational level and J designates the rotational level. By convention single primes refer to the upper state and double primes denote the lower state (see also Brittain et al. 2003). This relationship assumes that the angular extent of the emission source is equal to or greater than the beam size. However, even with the angular resolution of infrared spectroscopy (0″.7 beam), the hot emission source is considerably smaller than the area subtended by the beam. Correction for this diluted beam is discussed in § 4.3.

The column density in the v=2 state is related to the column densities of the individual rotational states by Boltzmann's equation,

$$N_{v=2,J'} = N_{v=2}(2J'+1)e^{-J'(J'+1)hcB/kT}/Q_r.$$

Here, $N_{v=2,J'}$ designates a rotational state in the vibrational level v=2, $N_{v=2}$ is the column density of the entire vibrational level v=2, (2J'+1) is the statistical weight of the transition, and $Q_r$ is the rotational partition function given by kT/hcB. To calculate the column density of CO in the state v=2, we plot the $\log(N_{v=2,J'})$ vs J' (J'+1)hcB/k such that the slope of the least squares fit is the inverse of the rotational temperature and the y-intercept is $\log(N_{v=2}/Q_r)$ (Fig. 4b). Thus the diluted column density of the v=2 vibrational state is $N_{v=2}=1.6\pm0.1 \times 10^{11}$ cm$^{-2}$.

If we assume the gas is in LTE, an expected condition for hot-dense gas in the inner disk (Najita et al. 2003), the column density of any level as well as the total column density of CO can be determined by

$$P_v = e^{-vhc\omega/kT}/Q_v,$$

where $P_v$ is the fractional population in state v, $\omega$ is the vibrational constant for CO (2170.21 cm$^{-1}$) and $Q_v$ is the vibrational partition function, given by $(1-e^{-hc\omega/kT})^{-1}$. The column density of CO in all states is then given by $N(CO)=N_{v=2}/P_{v=2}$. Since $P_{v=2}=1.3\pm0.3\times10^{-2}$, the diluted column density of the CO in all states is $N(CO)=1.2\pm0.3 \times 10^{13}$ cm$^{-2}$. For the vibrational level v=1, $P_{v=1}=0.11\pm0.1$ and $N_{v=1}=1.2\pm0.3 \times 10^{12}$.

### *4.3 Effects of Beam Dilution and Extinction*

The hot CO emission originates from a compact, unresolved region in the disk. To determine the physical column density from the CO emission lines, we must correct for both "beam dilution" and dust extinction at the wavelengths of the lines. Beam dilution plays a significant role in the interpretation of these data and highlights the importance of high angular resolution in the study of circumstellar disks. The angular extent of our beam was 0".7, which subtends an area of 8400 AU$^2$ at the distance of HL Tau (140 pc). The total physical column density is related to the diluted column density by:

$$N_{physical} = N_{diluted} e^{A_M/1.086} (8.4\times10^3/\pi r^2)$$

where $\pi r^2$ is the area subtended by the emitting CO and $A_M$ is the extinction in the M-band due to interstellar/circumstellar dust along the line of sight. It is the v=1 state that must remain optically thin after correcting for extinction and beam dilution. To correct for beam dilution, we assume the CO emission source subtends an area of 0.86 AU$^2$ if the disk is inclined 71±5° (see § 4.1). Thus the physical column density of CO in the state v=1 is $N_{v=1}= 1.2 \pm 0.3 \times 10^{16} e^{A_M/1.086}$.

We can now estimate $e^{A_M}$ by placing upper limits on the opacity in the v=2-1 emission lines. First, we consider the transition that is most optically thick at 1500 K (i.e., the transition that will show the largest deviation in the excitation plot). At 1500 K the most populated rotational level we observe is J=10, and it deviates from the linear fit presented in Figure 4b by 3σ when the line center opacity exceeds τ= 0.5. Next, we want to determine the column density of CO at 1500 K that gives the v=1, J=10 line an opacity of τ=0.5. To do this we perform a curve of growth analysis for a maximum opacity of

τ=0.5 and assume a Doppler line width of $b=3$ $km$ $s^{-1}$, which is a strict upper limit for 1500 K gas [3]. The maximum column density of optically thin CO in state v=1 at 1500 K is $N_{v=1} < 4 \times 10^{16} cm^{-2}$, so the 3σ upper limit on the M-band extinction is $A_M<1.3$ magnitudes. Even if the emitting area of the gas is underestimated by an order of magnitude, the M-band extinction is still less than 4 magnitudes. Thus our limit on the M-band extinction is consistent with the previous infrared extinction estimates of Close et al. (1997), Lucas & Roche (1998), and Lucas et al. (2004), all of which disagree with the 10 magnitudes of gray extinction proposed by Men'shchikov (1999).

## 5. DISCUSSION

The minimal infrared extinction we infer for this line of sight, $A_M < 1.3$, rules out a large population of gray particles along the line of sight, specifically, the 10 magnitudes of gray extinction suggested by Men'shchikov et al. (1999). The work by Lucas & Roche (1998) and Lucas et al. (2004) also suggests that Close et al. (1997) do not underestimate the extinction due to scattering. Furthermore, the minimal amount of $H_2O$ ice and silicates observed along the line of sight to HL Tau supports our contention that much of the dust is 'missing'. Whittet et al. (1988, 1989) showed that in most cases, embedded sources (where $A_V > 10$ magnitudes) revealed strong features from water ice, CO ice and silicates. While many factors can alter the relationship between extinction and these solid state features, it would be highly unusual to observe 52 magnitudes of extinction toward a star (as is implied by the observed CO gas) and not observe strong solid state features

Is reconciliation of the missing dust implied by the CO observations made possible by appealing to the inhomogenous physical structure of circumstellar disks? Particle settling in protoplanetary disks is a natural occurrence and causes the formation of higher density regions in the midplane (Chiang 2003, and references therein). Goldreich and Ward (1973) demonstrated that if sufficient density is reached in the midplane, gravitational instabilities lead directly to the formation of kilometer-sized bodies by direct assembly (directly from grains to kilometer-sized objects). The issue of turbulence has called this direct assembly approach into question (Weidenschilling 1995; see also Youdin & Shu 2002; Chiang 2004), but dust settling and grain growth are still expected. In a quiescent disk, Weidenschilling (1980) and Nakagawa et al. (1981, 1986) showed the settling of dust particles may be accelerated by particle coagulation and the sweep-up of smaller particles by larger ones due to size dependent settling velocities. A turbulent disk precedes the quiescent stage and quite naturally produces some particle growth (see Wiedenschilling 1984, Miyake and Nakagawa 1993). Observational quantification of gas/dust ratios will significantly improve our the understanding of dust settling and particle growth issues in the flared disks as well as constrain the midplane densities that in turn can directly impact planet formation scenarios.

---

[3] In the inner few tenths of an AU where the v=2-1 lines are formed, we need to know the velocity dispersion of the CO in state v=1. The sound speed of gas at 1500 K is ~ 5 km/s. Because the gas would be undergoing a continual shock and be dissipated in a very short time scale if it were supersonic, the turbulent velocity must be equal to or less than this value.

We cannot claim to have proved beyond a doubt that disk stratification is responsible for the anomalous gas to dust ratio of HL Tau. Our result, however, does provide an indication that dust settling may be a *measurable* process in circumstellar disks, and suggests that further studies could help to clarify its role in sculpting the evolution of disks and the formation of planets. For example, a highly varied sample of disks will constrain stratification in the physical structure of circumstellar disks and is the basis of ongoing research.

## 6. CONCLUSIONS

High-resolution infrared spectra of the narrow absorption and broad emission lines of gas phase CO molecules toward HL Tau constrain the basic physical properties of the inner accretion disk as well as the surrounding circumstellar envelope. The emission lines originate from the hot inner disk, typical of a young T Tauri star. The opacity of the v=2-1 emission lines constrain the M-band extinction to $A_M$ <1.3 magnitudes.

The narrow absorption lines from $^{12}CO$, $^{13}CO$, and $C^{18}O$ are found to originate at large distances from the central star, presumably where the line of sight intersects the flared portion of the disk. The temperature of the absorbing gas is ~100 K, and the composition of the CO isotopomers is consistent with interstellar values. The absorbing column of $^{12}CO$ ($7.5 \pm 0.2 \times 10^{18}$ cm$^{-2}$) indicates a large column of gas along the line of sight. The incongruence between the measured extinction and the extinction implied by this large column density of gas implies that there is a paucity of dust along the line of sight to HL Tau. Such a finding is consistent with the possibility that we are observing the vertical stratification of gas and dust in which the dust may have settled out of the envelope surrounding HL Tau and collapsed into the mid-plane of its circumstellar disk.

Acknowledgements: TWR and SDB were supported in part by NSF AST 02-05881. SDB was also supported in part under contract with the Jet Propulsion Laboratory (JPL) funded by NASA through the Michelson Fellowship Program. JPL is managed for NASA by the California Institute of Technology.

Table 1. Journal of Observations.

| Setting/Order | Spectral range (cm$^{-1}$) | Integration (min.) | S/N |
|---|---|---|---|
| MW1/16 | 2125-2158 | 4 | 80 |
| MW2/16 | 2127-2096 | 4 | 50 |
| K1/32 | 4188-4249 | 4 | 100 |

Table 2a. Rest position, observed position, radial velocity and equivalent width of $^{12}$CO absorption lines (v = 2-0 and v = 1-0).

| Line ID | $\tilde{\nu}_{rest}$ (cm$^{-1}$) | $\tilde{\nu}_{obs}$ (cm$^{-1}$) | $v_{rad}$ km s$^{-1}$ | W±δW (cm$^{-1}$) |
|---|---|---|---|---|
| (2,0) P(3) | 4248.32 | 4247.64 | 48 | 0.023±0.003 |
| (2,0) P(4) | 4244.26 | 4243.59 | 47 | 0.027±0.004 |
| (2,0) P(6) | 4235.95 | 4235.27 | 48 | 0.018±0.002 |
| (2,0) P(7) | 4231.69 | 4231 | 49 | 0.022±0.002 |
| (2,0) P(8) | 4227.35 | 4226.66 | 49 | 0.017±0.002 |
| (2,0) P(9) | 4222.95 | 4222.3 | 46 | 0.010±0.002 |
| (1,0) R(2) | 2154.6 | 2154.25 | 49 | 0.043±0.004 |
| (1,0) R(1) | 2150.86 | 2150.51 | 49 | 0.043±0.004 |
| (1,0) R(0) | 2147.08 | 2146.73 | 49 | 0.042±0.004 |
| (1,0) P(1) | 2139.43 | 2139.08 | 49 | 0.041±0.004 |
| (1,0) P(2) | 2135.55 | 2135.21 | 47 | 0.040±0.004 |
| (1,0) P(3) | 2131.63 | 2131.3 | 47 | 0.034±0.004 |
| (1,0) P(4) | 2127.68 | 2127.34 | 48 | 0.034±0.004 |
| (1,0) P(5) | 2123.7 | 2123.36 | 48 | 0.033±0.004 |
| (1,0) P(6) | 2119.68 | 2119.34 | 48 | 0.021±0.003 |
| (1,0) P(7) | 2115.63 | 2115.29 | 49 | 0.023±0.003 |
| (1,0) P(8) | 2111.54 | 2111.2 | 49 | 0.029±0.003 |
| (1,0) P(9) | 2107.42 | 2107.08 | 48 | 0.031±0.003 |
| (1,0) P(10) | 2103.27 | 2102.95 | 46 | <0.038±0.003 |
| (1,0) P(11) | 2099.08 | 2098.74 | 49 | 0.031±0.003 |

Table 2b. Rest position, observed position, radial velocity and equivalent width of $^{13}$CO absorption lines.

| line ID | $\tilde{\nu}_{rest}$ (cm$^{-1}$) | $\tilde{\nu}_{obs}$ (cm$^{-1}$) | $v_{rad}$ km s$^{-1}$ | W±δW (cm$^{-1}$) |
|---|---|---|---|---|
| (1,0) R(0) | 2099.71 | 2099.38 | 47 | 0.021±0.004 |
| (1,0) R(2) | 2106.9 | 2106.55 | 50 | 0.025±0.003 |
| (1,0) R(3) | 2110.44 | 2110.11 | 47 | 0.027±0.003 |
| (1,0) R(4) | 2113.95 | 2113.61 | 48 | 0.023±0.003 |
| (1,0) R(5) | 2117.43 | 2117.09 | 48 | 0.026±0.005 |
| (1,0) R(6) | 2120.87 | 2120.53 | 48 | 0.023±0.003 |
| (1,0) R(7) | 2124.29 | 2123.94 | 49 | 0.019±0.004 |
| (1,0) R(9) | 2131 | 2130.66 | 48 | 0.010±0.001 |
| (1,0) R(10) | 2134.31 | 2133.97 | 48 | 0.006±0.001 |
| (1,0) R(12) | 2140.83 | 2140.46 | 52 | 0.003±0.001 |

Table 2c. Rest position, observed position, radial velocity and equivalent width of C$^{18}$O absorption lines.

| line ID | $\tilde{\nu}_{rest}$ (cm$^{-1}$) | $\tilde{\nu}_{obs}$ (cm$^{-1}$) | $v_{rad}$ km s$^{-1}$ | W±δW (cm$^{-1}$) |
|---|---|---|---|---|
| (1,0) R(3) | 2106.44 | 2106.1 | 48 | 0.0068±0.0013 |
| (1,0) R(4) | 2109.94 | 2109.6 | 48 | 0.0074±0.0023 |
| (1,0) R(5) | 2113.41 | 2113.06 | 49 | 0.0074±0.0017 |
| (1,0) R(6) | 2116.84 | 2116.5 | 48 | 0.0041±0.0009 |
| (1,0) R(7) | 2120.24 | 2119.89 | 49 | 0.0040±0.0028 |

Table 3. Rest position, observed position, radial velocity and measured flux for $^{12}$CO emission lines.

| Line ID | $\tilde{\nu}_{rest}$ (cm$^{-1}$) | $\tilde{\nu}_{obs}$ (cm$^{-1}$) | $v_{rad}$ km s$^{-1}$ | F±δ(F) (10$^{-16}$ W m$^{-2}$) |
|---|---|---|---|---|
| (1,0) R(2) | 2154.6 | 2154.37 | 32 | 1.00±.06 |
| (1,0) R(1) | 2150.86 | 2150.49 | 52 | 0.94±.06 |
| (1,0) R(0) | 2147.08 | 2146.9 | 25 | 0.64±.07 |
| (1,0) P(1) | 2139.43 | 2139.03 | 56 | 0.76±.07 |
| (1,0) P(2) | 2135.55 | 2135.27 | 39 | 0.98±.07 |
| (1,0) P(3) | 2131.63 | 2131.32 | 44 | 1.20±.07 |
| (1,0) P(4) | 2127.68 | 2127.43 | 35 | 1.20±.07 |
| (1,0) P(5) | 2123.7 | 2123.38 | 45 | 0.95±.05 |
| (1,0) P(6) | 2119.68 | 2119.28 | 57 | 0.87±.05 |
| (1,0) P(7) | 2115.63 | 2115.25 | 54 | 0.87±.05 |
| (1,0) P(8) | 2111.54 | 2111.19 | 50 | 0.83±.06 |
| (1,0) P(9) | 2107.42 | 2107 | 60 | 0.69±.06 |
| (1,0) P(10) | 2103.27 | 2102.97 | 43 | 0.80±.06 |
| (1,0) P(11) | 2099.08 | 2098.74 | 49 | 0.86±.06 |
| (1,0) P(28) | 2022.91 | 2022.62 | 43 | 1.20±.08 |
| (1,0) P(30) | 2013.35 | 2013.01 | 51 | 0.99±.08 |
| (1,0) P(31) | 2008.53 | 2008.18 | 52 | 1.20±.08 |
| (1,0) P(32) | 2003.67 | 2003.37 | 45 | 1.10±.08 |
| (1,0) P(36) | 1983.94 | 1983.57 | 56 | 0.52±.08 |
| (1,0) P(37) | 1978.93 | 1978.61 | 49 | 0.52±.08 |
| (1,0) P(38) | 1973.89 | 1973.44 | 68 | 0.46±.08 |
| (2,1) R(10) | 2156.36 | 2156.1 | 36 | 0.55±0.05 |
| (2,1) R(9) | 2152.94 | 2152.68 | 36 | 0.54±0.05 |
| (2,1) R(7) | 2146 | 2145.66 | 48 | 0.57±0.05 |
| (2,1) R(6) | 2142.47 | 2142.24 | 32 | 0.47±0.05 |
| (2,1) P(23) | 2020.6 | 2020.34 | 39 | 0.46±0.04 |
| (2,1) P(25) | 2011.42 | 2011.15 | 40 | 0.45±0.04 |
| (2,1) P(26) | 2006.78 | 2006.5 | 42 | 0.42±0.04 |
| (2,1) P(27) | 2002.12 | 2001.77 | 52 | 0.38±0.04 |
| (2,1) P(28) | 1997.42 | 1997.06 | 54 | 0.35±0.04 |
| (2,1) P(30) | 1987.92 | 1987.76 | 24 | 0.28±0.04 |
| (2,1) P(31) | 1983.13 | 1982.76 | 56 | 0.27±0.04 |
| (2,1) P(32) | 1978.31 | 1978.07 | 36 | 0.31±0.04 |

Table 4. Excitation Temperatures, Column Densities, and Optical Depths Derived from CO Absorption Lines

| Ro-vibrational band | Rotational Temperature (K) | Column Density (cm$^{-2}$) |
|---|---|---|
| $^{12}CO(2,0)$ | $105 \pm 8$ | $(7.5 \pm 0.2) \times 10^{18}$ |
| $^{12}CO(1,0)^a$ | ~100K | $7 \times 10^{18}$ |
| $^{13}CO(1,0)$ | $80 \pm 10$ | $(9.9 \pm 1.1) \times 10^{16}$ |
| $C^{18}O(1,0)$ | $80 \pm 20$ | $(9 \pm 2) \times 10^{15}$ |

$^a$ Loosely fit to match the (2,0) band.

Figure 1a-g CO spectrum from HL Tau. The broad emission features result from hot CO gas near the star. The narrow absorption features that are superimposed on these emission features have a lower rotational temperature indicating colder gas along the line of sight. The strength of the overtone lines of $^{12}CO$, relative to the fundamentals, suggests that the fundamental lines are optically thick. The fact that the optically thick lines do not *appear* saturated means their intrinsic width is much narrower than the resolution of NIRSPEC (12 km/s).

Figure 2 Curve of Growth. We have corrected for the saturation of the absorption lines using a curve of growth analysis. By plotting log(W/ṽ) vs log (Nfṽ), where f is the oscillator strength, we relate the measured equivalent widths to the column densities by deriving the effects of opacity upon the Gaussian line profile. This plot demonstrates the consistency of all the data with a common velocity dispersion (*b=1.3±0.1 km/s*).

Figure 3 The rotational temperature of the CO absorption lines. The $\log_{10}(N_J/(2J+1))$ vs $E_J/k$ is plotted for each isotopomer such that the negative reciprocal of the slope is proportional to the rotational temperature. The plots are linear because they have been corrected for opacity. The uncertainty in the temperature only includes the uncertainty in the fit. It does not account for the uncertainty in *b*; however, the temperatures are consistent within 3σ. The cool temperatures indicate the gas is from the outer parts of the disk/envelope.

Figure 4ab The excitation plot of the CO ro-vibrational emission lines. A straight line fit reveals optically thin LTE gas such that the negative reciprocal of the slope is proportional to the rotational temperature of the gas, and the y-intercept is related to the total column density by, $b=\log(N_{v=2}/Q_r)$. 4a) The fundamental band (v=1-0), shows the CO lines cannot be fit with a straight line which implies that the gas is optically thick. 4b) The fit to the hot band lines (v=2-1) is linear (optically thin) and implies that the rotational temperature of the hot CO is 1500±100 K.

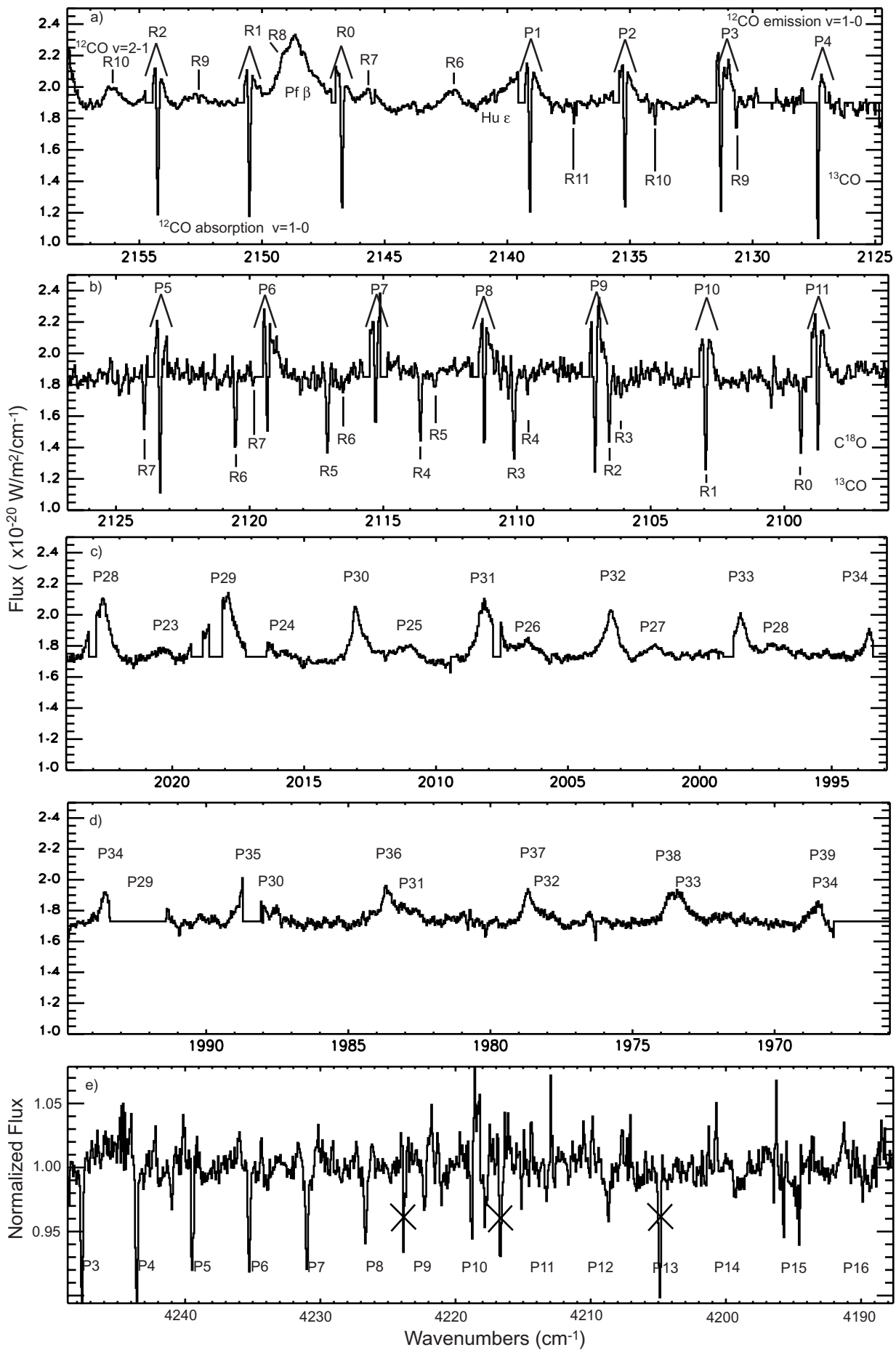

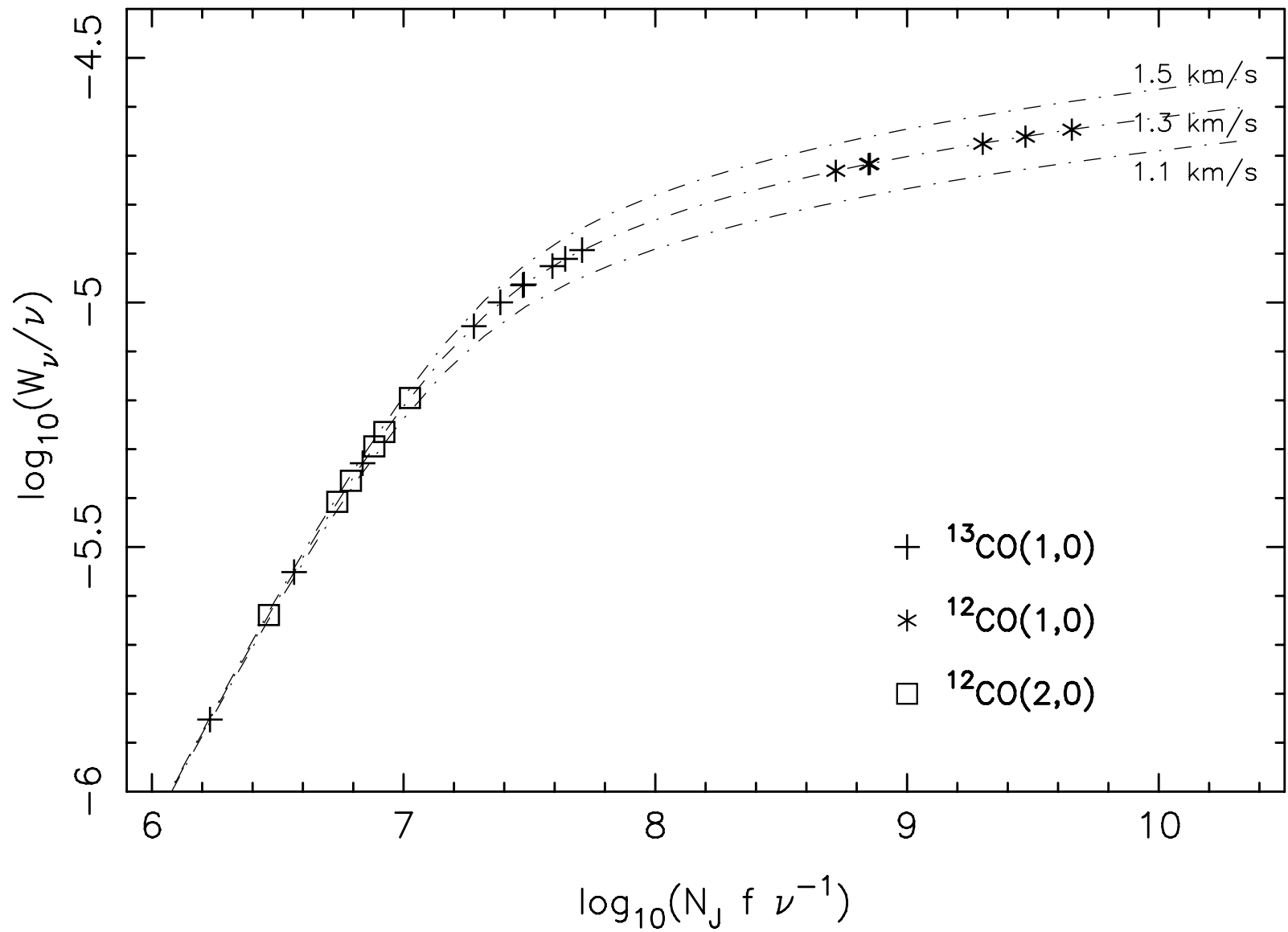

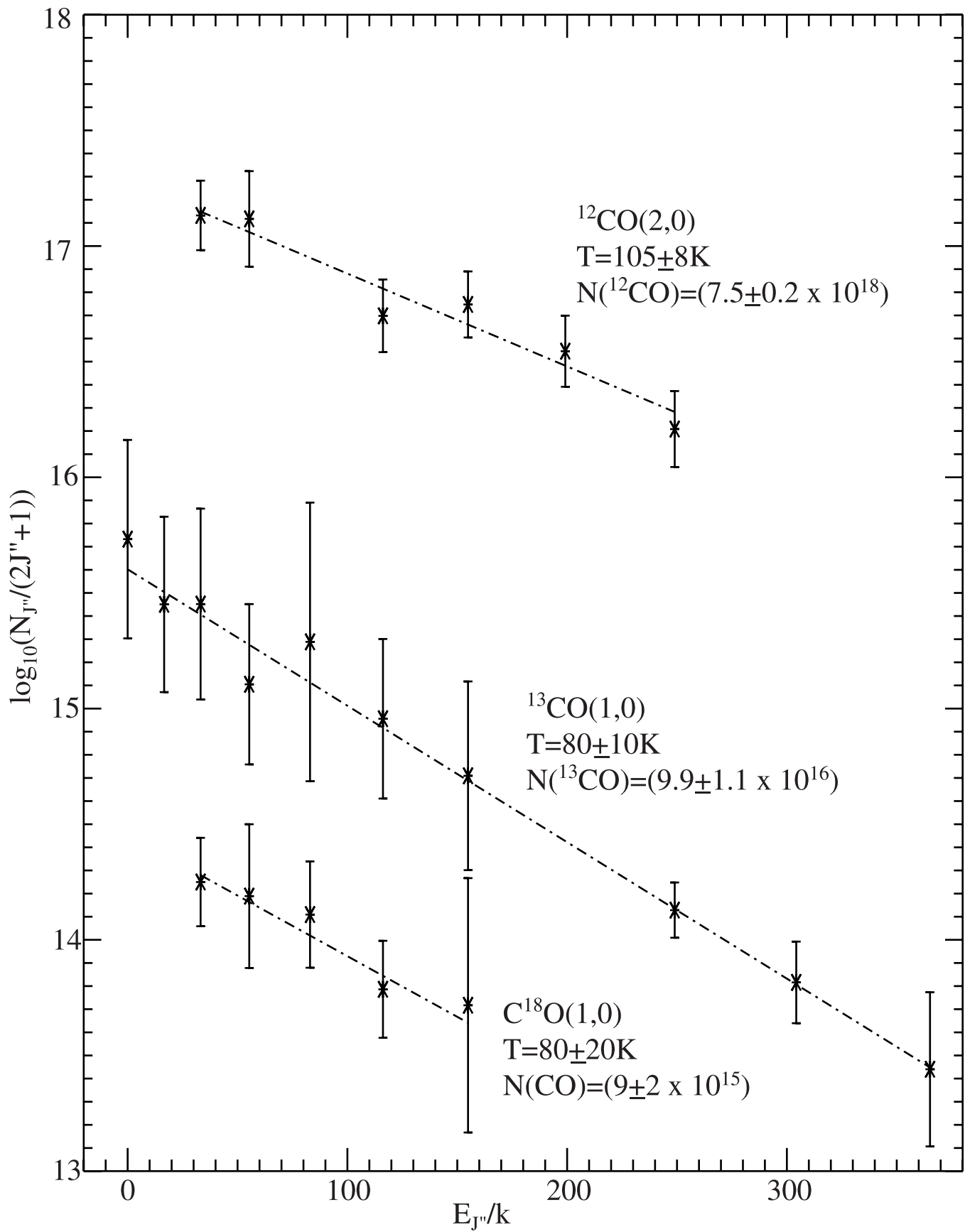

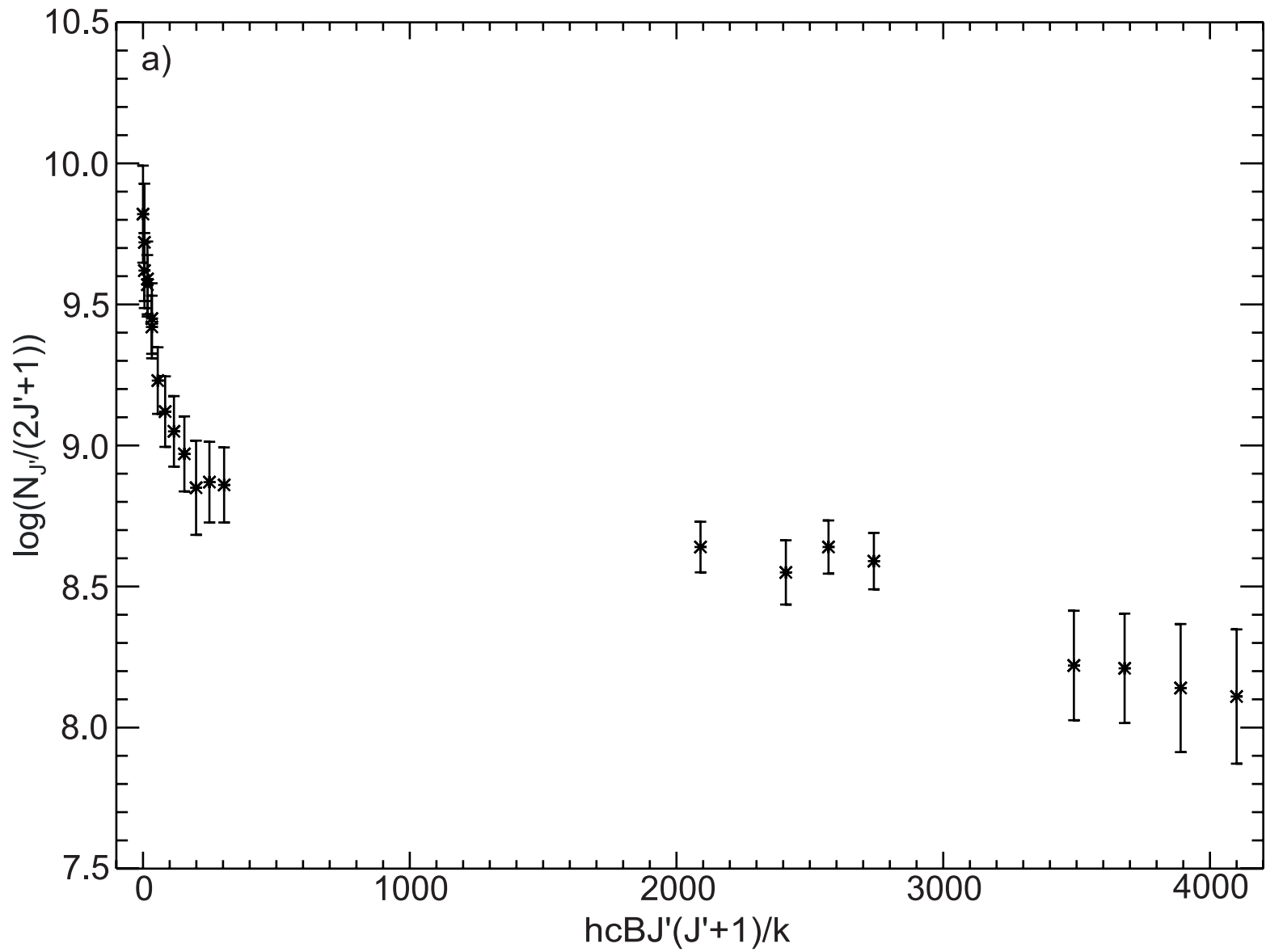

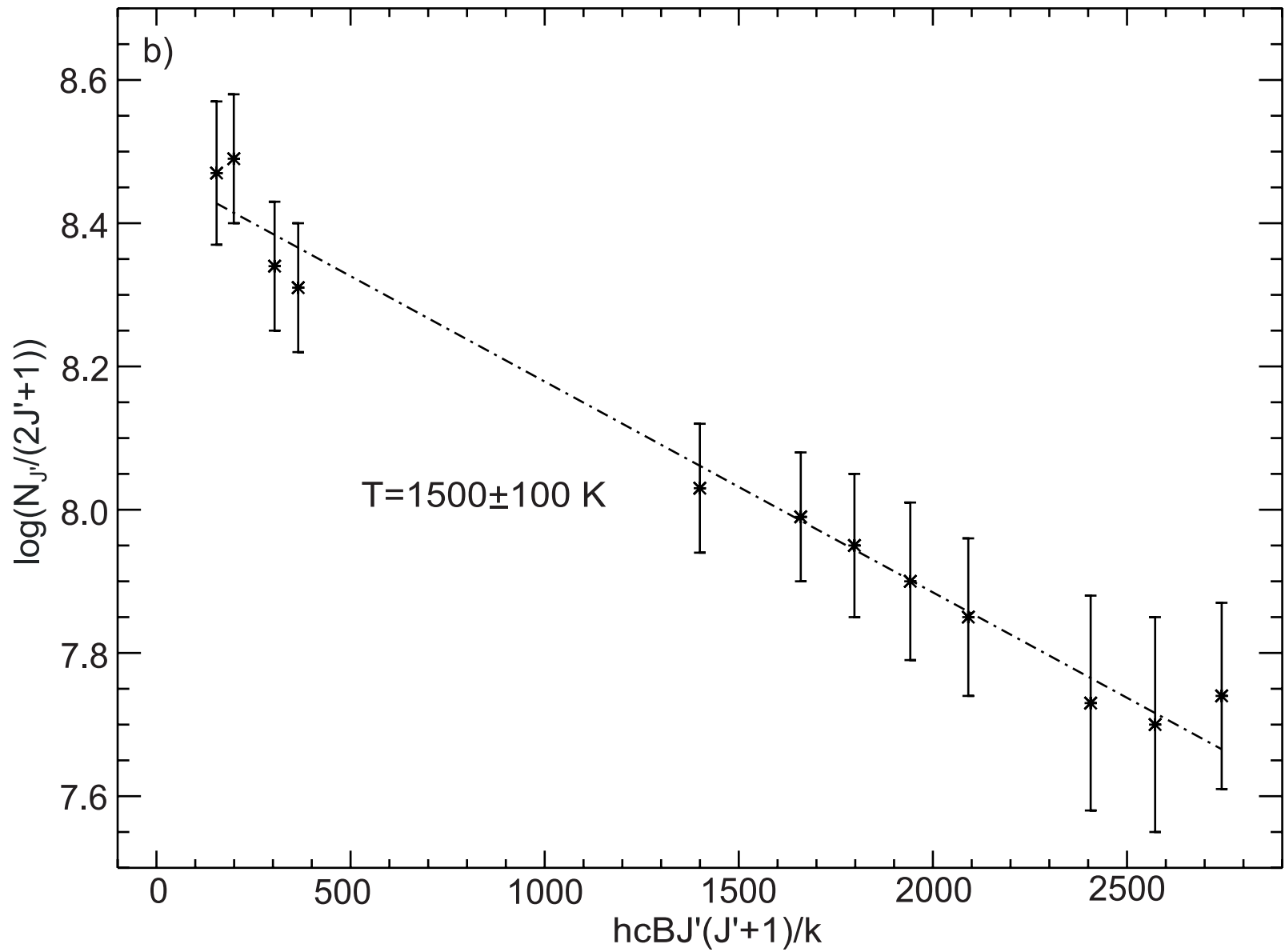